\documentclass[prl,twocolumn]{revtex4}
\usepackage{graphicx}

\begin{document}
\title{Preparation of a quantum state with one molecule at each site of an optical lattice}
\author{T. Volz, N. Syassen, D.~M. Bauer, E. Hansis, S. D{\"u}rr, and G. Rempe}
\affiliation{Max-Planck-Institut f{\"u}r Quantenoptik, Hans-Kopfermann-Stra{\ss}e 1, 85748 Garching, Germany}
\hyphenation{Fesh-bach}

\maketitle

\bf

Ultracold gases in optical lattices are of great interest, because these systems bear a great potential for applications in quantum simulations and quantum information processing, in particular when using particles with a long-range dipole-dipole interaction, such as polar molecules \cite{demille:02,goral:02,lee:05,micheli:06,barnett:06}. Here we show the preparation of a quantum state with exactly one molecule at each site of an optical lattice. The molecules are produced from an atomic Mott insulator \cite{greiner:02} with a density profile chosen such that the central region of the gas contains two atoms per lattice site. A Feshbach resonance is used to associate the atom pairs to molecules \cite{regal:03,herbig:03,duerr:04,strecker:03,cubizolles:03,jochim:03a,xu:03,koehler:06}. Remaining atoms can be removed with blast light \cite{xu:03,thalhammer:06}. The technique does not rely on the molecule-molecule interaction properties and is therefore applicable to many systems.

\rm

A variety of interesting proposals for quantum information processing and quantum simulations \cite{demille:02,goral:02,lee:05,micheli:06,barnett:06} require as a prerequisite a quantum state of ultracold polar molecules in an optical lattice, where each lattice site is occupied by exactly one molecule. A promising strategy for the creation of such molecules is based on association of ultracold atoms using a Feshbach resonance or photoassociation and subsequent transfer to a much lower ro-vibrational level using Raman transitions \cite{sage:05}. If the molecule-molecule interactions are predominantly elastic and effectively repulsive, then a state with one molecule per lattice site can finally be obtained using a quantum phase transition from a superfluid to a Mott insulator by ramping up the depth of an optical lattice \cite{greiner:02}. However, many molecular species do not have such convenient interaction properties, so that alternative strategies are needed. Here, we demonstrate a technique that is independent of the molecule-molecule interaction properties. The technique relies on first forming an atomic Mott insulator and then associating molecules. Several previous experiments \cite{thalhammer:06,rom:04,stoferle:06,winkler:06,ryu:cond-mat/0508201} associated molecules in an optical lattice, but none of them demonstrated the production of a quantum state with exactly one molecule per lattice site. Another interesting perspective of the state prepared here is that after Raman transitions to the ro-vibrational ground state, one could lower the lattice potential and obtain a Bose-Einstein condensate (BEC) of molecules in the ro-vibrational ground-state \cite{jaksch:02,damski:03}.

The behavior of bosons in an optical lattice is described by the Bose-Hubbard Hamiltonian \cite{jaksch:98}. The relevant parameters are the amplitude $J$ for tunneling between neighboring lattice sites and the on-site interaction matrix element $U$. We create a Mott insulator \cite{greiner:02} of atomic $^{87}$Rb starting from an atomic BEC in an optical dipole trap by slowly ramping up the depth of the optical lattice (see Methods). The typical lattice depth \cite{greiner:02} seen by an atom is $V_0=24 E_r$, where $E_r=\hbar^2k^2/(2m)$ is the recoil energy, $m$ is the mass of one atom, and $2\pi/k=830$~nm is the wavelength of the lattice light. At this lattice depth, the atomic tunneling amplitude is $J=2\pi\hbar\times 4$~Hz.

\begin{figure}[tb]
\includegraphics[width=.45\textwidth]{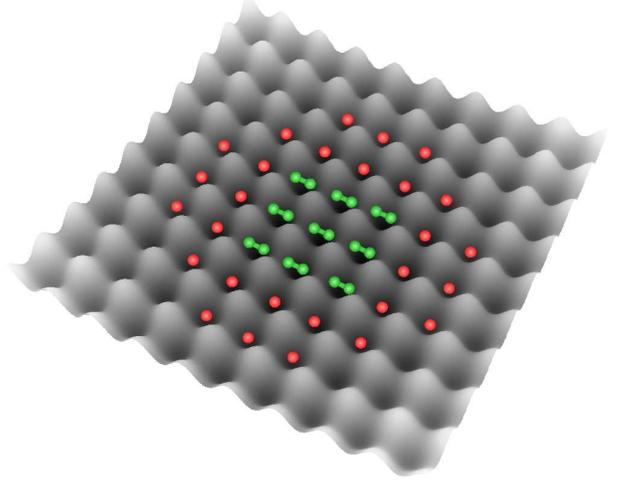}
\caption{\label{fig-scheme}
{\bf Scheme of the molecular $n=1$ state.} In the core of the cloud, each lattice site is occupied by exactly $n=1$ molecule (shown in green). In the surrounding shell, each site is occupied by exactly one atom (shown in red). The atoms can be removed with a blast laser. In the experiment, the number of occupied lattice sites is much larger than shown here.
 }
\end{figure}

Due to the external harmonic confinement (see Methods), the atomic Mott insulator is inhomogeneous. It consists of shells of constant lattice filling with exactly $n$ particles per lattice site. Neighboring shells are connected by narrow superfluid regions. For a deep lattice, $J$ becomes small. With the approximation $J=0$, the ground state of the system including the harmonic confinement can be calculated analytically. This model predicts that the fraction of atoms occupying sites with $n=2$ atoms has a maximum of 53\%. This value should be reached, when operating close to the point where a core with $n=3$ atoms per site starts to form. In order to operate at this point, we load the lattice with $N=10^5$ atoms. For larger atom number, an $n=3$ core forms, which is seen in the experiment as loss when associating molecules (see below).

After preparing the atomic Mott insulator at a magnetic field of $B=1008.8$~G, molecules are associated as described in Ref.~\cite{duerr:04}. To this end, the magnetic field is slowly (at 2~G/ms) ramped across the Feshbach resonance at 1007.4~G \cite{marte:02} to a final value of $B=1006.6$~G. At lattice sites with a filling of $n=1$, this has no effect. At sites with $n>1$, atom pairs are associated to molecules. If the site contained $n>2$ atoms, then the molecule can collide with other atoms or molecules at the same lattice site. As the molecules are associated in a highly-excited ro-vibrational state, the collisions are likely to be inelastic. This leads to fast loss of the molecule and its collision partner from the trap. The association ramp lasts long enough to essentially empty all sites with $n>2$ atoms. For lattice sites with $n=2$ atoms, the association efficiency is above 80\%, similar to Ref.~\cite{thalhammer:06}. The resulting molecular $n=1$ state is sketched in Fig.~\ref{fig-scheme}. The maximum fraction of the population that can be converted into molecules (measured as the atom number reappearing after dissociation) is found to be 47(3)\%, which is close to the theoretical limit of 53\% discussed above.

\begin{figure}[tb]
\includegraphics[width=.45\textwidth]{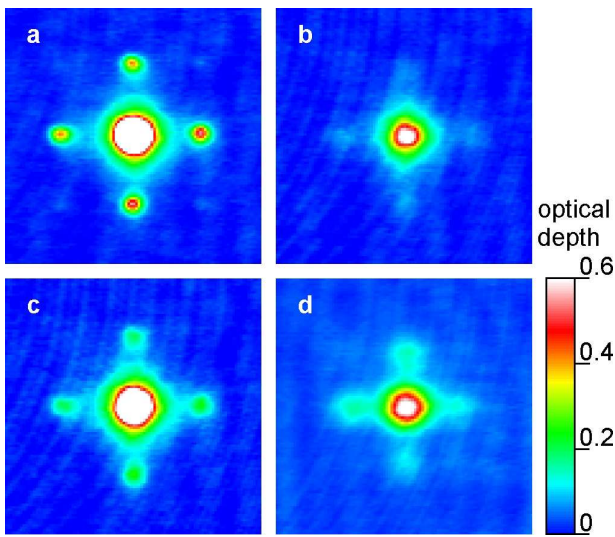}
\caption{\label{fig-restoration}
{\bf Atomic Mott insulator and molecular $n=1$ state.}
{\bf a} An atomic Mott insulator is melted by reducing the lattice depth slowly. The system returns to the superfluid phase and phase coherence is restored. This phase coherence is probed by quickly switching the lattice off and observing an atomic interference pattern in time of flight.
{\bf b} After association of molecules, only lattice sites occupied by $n=1$ atoms contribute to the signal.
{\bf c} After association and dissociation of molecules, the satellite peaks are much stronger than in {\bf b}, thus proving that the molecular part of the cloud was in a molecular $n=1$ state.
{\bf d.} Pure molecular $n=1$ state. Same as {\bf c} but between association and dissociation, remaining atoms were removed with blast light. The visibilities in {\bf a-d} are 0.93(2), 0.80(5), 0.86(1), and 0.61(2), calculated from squares with areas corresponding to atomic momenta of $0.22 \hbar k \times 0.22 \hbar k$ (see inset in Fig.~\ref{fig-satellite-lifetime} and Ref.~\cite{gerbier:05}).
 }
\end{figure}

At a lattice depth of $V_0=24E_r$ for atoms, the tunneling amplitude for molecules is calculated to be $J_m=2\pi\hbar \times 0.3$~mHz. This is negligible compared to the hold time between association and dissociation, so that the positions of the molecules are frozen. This conclusion is further supported by the experimental observation of a long lifetime of the molecules (see below).

In order to show that the molecular part of the sample really is in the $n=1$ state, the molecules are first dissociated back into atom pairs by slowly (at 1.5~G/ms) ramping the magnetic field back across the Feshbach resonance. This brings the system back into the atomic Mott insulator state with shells with $n=1$ and $n=2$. Then, the atomic Mott insulator is melted by slowly (within 10~ms) ramping down the lattice from $V_0=24 E_r$ to $V_0=4 E_r$. Finally, the lattice is quickly switched off and after some time of flight an absorption image is taken.

Such images are shown in Fig.~\ref{fig-restoration}. Part {\bf a} shows the result if the magnetic-field ramp for association and dissociation of molecules is omitted. This matter-wave interference pattern shows that phase coherence is restored when ramping down the lattice, thus demonstrating that an atomic Mott insulator is realized at 1008.8~G. Part {\bf b} shows the pattern obtained if molecules are associated but not dissociated, so that they remain invisible in the detection. This signal comes only from sites with $n=1$ atoms. Part {\bf c} shows the result obtained for the full sequence with association and dissociation of molecules. Obviously, the satellite peaks regain considerable population compared to {\bf b}, which proofs that after dissociation, we recover an atomic Mott insulator. This shows that association and dissociation must have been coherent and adiabatic. Combined with the freezing of the positions of the molecules and the fact that the association starts from an atomic Mott insulator with an $n=2$ core, this implies that the molecular part of the cloud must have been in a quantum state with one molecule per lattice site.

After associating the molecules, remaining atoms can be removed from the trap using microwave radiation and a blast laser as in Ref.~\cite{thalhammer:06}. This produces a pure molecular sample. The molecule numbers before and after the blast are identical within the experimental uncertainty of 5\%. In order to show that the pure molecular sample is in the $n=1$ state, the molecules are dissociated, the lattice is ramped down to $V_0=1.2 E_r$ within 30~ms, ramped back up (see Methods) to $V_0=6 E_r$ within 5~ms, and finally switched off. The result is shown in Fig.~\ref{fig-restoration}{\bf d}. Again, an interference pattern is visible. The time of flight was 12~ms in {\bf a}-{\bf c} and 11~ms in {\bf d}.

\begin{figure}[tb]
\includegraphics[width=.45\textwidth]{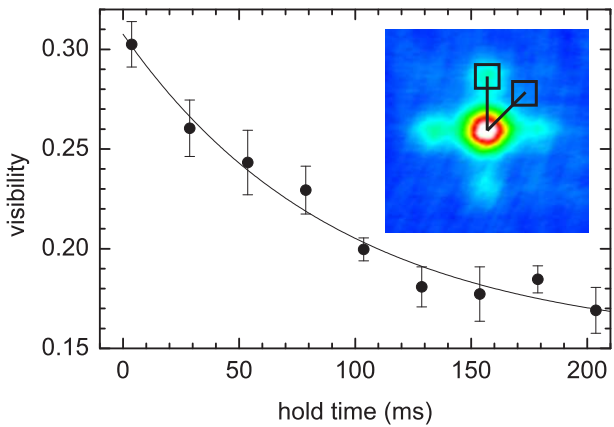}
\caption{\label{fig-satellite-lifetime}
{\bf Lifetime of the molecular $n=1$ state after removing the atoms.} The visibility of the satellite peaks in measurements similar to Fig.~\ref{fig-restoration}{\bf d} decays as a function of the hold time between molecule association and dissociation. The visibility was calculated from squares (as shown in the inset) with areas corresponding to atomic momenta of $1.0\hbar k\times1.0\hbar k$. The line shows an exponential fit which yields a $1/e$-lifetime of 93(22)~ms. Error bars are statistical (one standard deviation).
 }
\end{figure}

The height of the satellite peaks can be quantified using the visibility defined in Ref.~\cite{gerbier:05}. Fig.~\ref{fig-satellite-lifetime} shows the decay of the visibility as a function of the hold time between molecule association and dissociation. These data were obtained from measurements as in Fig.~\ref{fig-restoration}{\bf d} except that after dissociation the lattice was ramped down to $V_0=2.8 E_r$ within 10~ms and ramped back up to $V_0=5.5 E_r$ within 4~ms. The observed lifetime of the visibility is sufficient for many applications. For comparison, the measured lifetime of the molecule number is 160(20)~ms, which is probably due to scattering of lattice photons \cite{thalhammer:06}. This molecule loss sets an upper limit on the lifetime of the visibility, because the sites at which molecules are lost are randomly distributed across the lattice, thus gradually destroying the molecular $n=1$ state. According to the fit, the visibility settles to an offset value of 15\%. This might be partly due to the fact that the Wannier function \cite{gerbier:05} for small lattice depth is not spherically symmetric and partly due to a small contribution of non-removed atoms because of imperfections of the blast laser (15\% of the total signal at zero hold time).

\begin{figure}[tb]
\includegraphics[width=.45\textwidth]{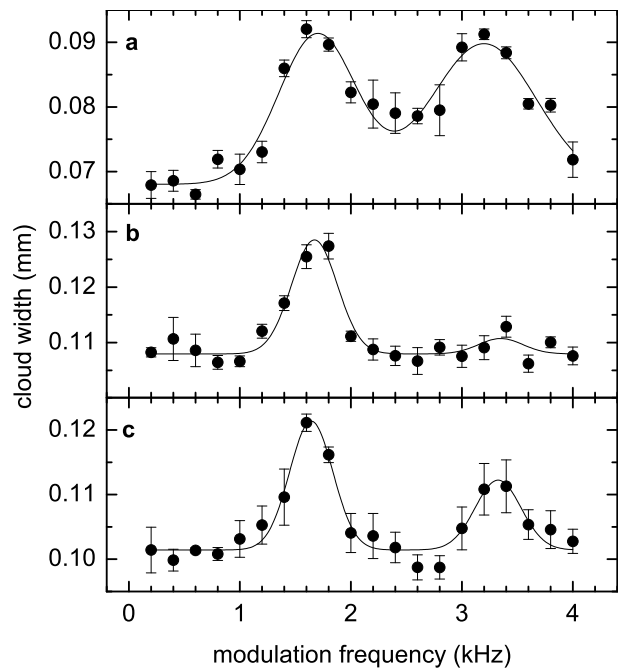}
\caption{\label{fig-excitation-spectrum}
{\bf Excitation spectrum of the atomic Mott insulator.} The full width at half maximum of the central interference peak is shown as a function of the frequency at which the lattice depth is modulated. The results in parts {\bf a-c} correspond to the conditions of Fig.~\ref{fig-restoration}{\bf a-c}. First, the usual lattice ramp-down starting at $V_0=24 E_r$ is interrupted at $V_0=15 E_r$. Next, the power of one lattice beam is modulated for 11~ms with a peak-to-peak amplitude of 50\%. Finally, the lattice ramp-down continues as usual. Resonances are visible at 1.6 and 3.2~kHz. The lines are a guide to the eye. Error bars are statistical.
}
\end{figure}

Figure~\ref{fig-excitation-spectrum} shows the excitation spectrum of the atomic Mott insulator at $V_0=15 E_r$ ($J=2\pi\hbar \times 21$~Hz, $J_m=2\pi\hbar \times 14$~mHz) as measured by amplitude modulation of one lattice beam \cite{stoeferle:04}. The spectrum in Fig.~\ref{fig-excitation-spectrum}{\bf a} shows clear resonances at energies $U$ and $2U$. Below the first resonance, no noticeable excitations are created, showing that the excitation spectrum has a gap. This again shows that the system before molecule association is an atomic Mott insulator. In Fig.~\ref{fig-excitation-spectrum}{\bf b}, the signal at $2U$ essentially disappeared, because the signal at $2U$ is created by processes that require lattice sites with $n\geq2$ atoms \cite{greiner:02}, which are absent without the modulation. The spectrum in Fig.~\ref{fig-excitation-spectrum}{\bf c} is similar to the one in Fig.~\ref{fig-excitation-spectrum}{\bf a}. Resonances at $U$ and $2U$ are clearly visible in Fig.~\ref{fig-excitation-spectrum}{\bf c}. This gives further experimental support for the above conclusion that the system after the association-dissociation ramp is still an atomic Mott insulator.

We also measured the excitation spectrum at various lattice depths for the molecular $n=1$ state after removing the atoms, corresponding to Fig.~\ref{fig-restoration}{\bf d}. This spectrum does not show any resonances related to $U$. We only observe resonances at much higher frequencies due to band excitation. The absence of resonances related to $U$ is probably due to the short lifetime of two molecules at one lattice site (see Methods), which leads to an estimated resonance width of $\Gamma=2\pi\hbar \times 10$~kHz. This is too broad and consequently too shallow to be observed. Furthermore, the real part of the molecule-molecule scattering length is unknown. Hence, it is also unknown whether the molecular $n=1$ state created here has a gap, and if so, at what energy the first resonance should be expected.

In the experiment $\Gamma\gg J_m$, so that the effective tunneling rate between neighboring lattice sites is $4J_m^2/(\hbar\Gamma)$ \cite{cohen-tannoudji:92:p49:ver2}. Interestingly, fast on-site decay $\Gamma$ suppresses the mobility in the many-body system. This might lead to an insulator-like behavior without a gap.

In conclusion, we prepared a quantum state with exactly one molecule at each site of an optical lattice. It is an interesting question, whether gapless systems can have insulating properties due to fast on-site losses. The quantum state prepared here is exactly the state that a molecular Mott insulator has in the limit of negligible tunneling. Unlike the creation of a molecular Mott state by a quantum phase transition from a molecular BEC, our method works independently of the molecule-molecule interaction properties.

\bigskip
{\bf METHODS}

{\bf Dipole trap setup}

The experiment begins with the creation of a BEC of $^{87}$Rb atoms in a magnetic trap \cite{marte:02}. Once created, the BEC is transferred into an optical dipole trap which is created by crossing two light beams at right angles. One beam has a wavelength of 1064~nm, a power of 170~mW, and a waist ($1/e^2$ radius of intensity) of 140~$\mu$m. The other beam has a wavelength of 1050~nm, a power of 2.1~W, and an elliptically shaped focus with waists of 60~$\mu$m and 700~$\mu$m. The crossed-beam dipole trap creates an approximately harmonic confinement with measured trap frequencies of 20, 20, and 110~Hz. The strongest confinement supports the atoms against gravity. After loading the dipole trap, a magnetic field of approximately 1000~G is applied. The atoms are transferred \cite{volz:03} to the absolute ground state $|F=1,m_F=1\rangle$, which has a Feshbach resonance at 1007.4~G with a width of 0.2~G \cite{volz:03,duerr:04a}. The dipole-trap light at 1050~nm comes from a multi-frequency fiber laser with a linewidth of $\sim 1$~nm. The experiment shows that this causes fast loss of the molecules, presumably due to photodissociation. Therefore, the power of the dipole-trap light is slowly reduced to zero just before associating the molecules and slowly ramped back up to its original value just after dissociating them. Finally, the dipole trap and the optical lattice are switched off simultaneously.

{\bf Lattice setup}

A simple-cubic optical lattice is created by illuminating the BEC with three retro-reflected light beams with a waist of 140~$\mu$m. The finite waists of the lattice beams give rise to an additional overall confinement. For $V_0=15 E_r$, this corresponds to an estimated trap frequency of 50~Hz. This harmonic potential adds to the potential of the dipole trap. Note that the polarizability of a Feshbach molecule is approximately twice as large as for one atom. Hence, the molecules experience a lattice depth of $2V_0$. 

{\bf Ramping for conditions used in Fig.~\ref{fig-restoration}{\bf d}}

To understand why a special ramping procedure is needed for Fig.~\ref{fig-restoration}{\bf d}, consider the atomic Mott insulator obtained after dissociating the molecules. This Mott insulator has only the $n=2$ core, while the surrounding $n=1$ shell is missing. Now consider two atoms at an $n=2$ site. If one of the atoms tunneled to an empty neighboring lattice site, then this would release an energy $U$. But there is no reservoir that could absorb this energy so that the tunneling is suppressed \cite{winkler:06}. Hence, a strong reduction of $U$ is required to melt the pure $n=2$ Mott insulator. This can be achieved by ramping the lattice down to a point much below $V_0=4 E_r$ (or by reducing the scattering length). At this low lattice depth, the sudden switch-off of the lattice does not produce noticeable satellite peaks. To observe such peaks, the lattice therefore must be ramped back up before switching it off.

{\bf Inelastic collisions}

If one considers only lattice sites which are occupied by exactly two molecules, and if tunneling is negligible, then the sites will decay independently of each other. Hence, the total population of these sites will decay exponentially. We measured the rate coefficient for inelastic molecule-molecule collisions. At $V_0=15 \; E_r$, this leads to an estimated lifetime of $16~\mu$s.

\bigskip

{\bf Acknowledgements}

We thank J.~I. Cirac, T. Esslin\-ger, and W. Zwerger for fruitful discussions.

\end{document}